\documentclass[conference,a4paper]{IEEEtran}
\usepackage{cite,graphicx,multirow,multicol,setspace,amsmath,wrapfig}
\usepackage{tabulary,amssymb,adjustbox,booktabs,qtree,tikz-qtree,synttree}
\usepackage[flushleft]{threeparttable}
\usepackage{makecell,eqnarray}
\usepackage{cuted}
\usepackage[linesnumbered,ruled]{algorithm2e}

\graphicspath{ {Images/} }

\makeatletter
\let\MYcaption\@makecaption
\makeatother

\usepackage[font=footnotesize]{subcaption}

\makeatletter
\let\@makecaption\MYcaption
\makeatother

\begin{document}

\title {Structure Selection of Polynomial NARX Models using Two Dimensional (2D) Particle Swarms}

\author{\IEEEauthorblockN{Faizal Hafiz\IEEEauthorrefmark{1},
Akshya Swain\IEEEauthorrefmark{1},
Eduardo MAM Mendes\IEEEauthorrefmark{2} and
Nitish Patel\IEEEauthorrefmark{1}}

\IEEEauthorblockA{\IEEEauthorrefmark{1}Department of Electrical \& Computer Engineering,
The University of Auckland, Auckland, New Zealand.\\
Email: faizalhafiz@ieee.org, a.swain@auckland.ac.nz}

\IEEEauthorblockA{\IEEEauthorrefmark{2}Department of Electronics Engineering, Federal University of Minas Gerais, Belo Horizonte, Brazil.\\
}
}
\maketitle

\begin{abstract}

The present study applies a novel two-dimensional learning framework (2D-UPSO) based on particle swarms for structure selection of polynomial nonlinear auto-regressive with exogenous inputs (NARX) models. This learning approach explicitly incorporates the information about the \textit{cardinality} (\textit{i.e.}, the number of terms) into the structure selection process. Initially, the effectiveness of the proposed approach was compared against the classical genetic algorithm (GA) based approach and it was demonstrated that the 2D-UPSO is superior. Further, since the performance of any meta-heuristic search algorithm is critically dependent on the choice of the fitness function, the efficacy of the proposed approach was investigated using two distinct information theoretic criteria such as Akaike and Bayesian information criterion. The robustness of this approach against various levels of measurement noise is also studied. Simulation results on various nonlinear systems demonstrate that the proposed algorithm could accurately determine the structure of the polynomial NARX model even under the influence of measurement noise.          

\end{abstract}


\IEEEpeerreviewmaketitle

\section{Introduction}
System identification addresses the issue of constructing mathematical models from the observed input-output data and has been a major research concern from diverse fields such as statistics, control theory, information theory, economics, ecology, agriculture and others~\cite{Billings:2013,Ljung:1999}. Although, methods for linear systems identification are now well established, the development of nonlinear system identification methods, which can be valid for a broad class of nonlinear systems, is a research issue. This is partly due to the well-recognized highly individualistic nature of nonlinear systems which restricts the unifying dynamical features that are amenable to system identification.

The first step in the nonlinear system identification is the choice of the model amongst various models which are used to represent the system such as Volterra, Wiener, neural networks, polynomial models, rational models and others. The focus of this study is the identification of nonlinear systems represented by polynomial nonlinear auto-regressive with exogenous inputs (NARX) models~\cite{Billings:2013}. Although this model could represent a wide class of nonlinear systems, the number of possible terms increases exponentially with the increase in the order of non-linearity and maximum lags of inputs and outputs. Inclusion of all the terms is not desirable as the parameter estimation problem may become ill-conditioned for the consequent complex model~\cite{Billings:2013}. Therefore, determination of the model structure or which terms to include in the model is essential if a parsimonious model is to be determined from the large number of candidate terms. 

During the past three decades, researchers have proposed several algorithms based on orthogonal least squares, evolutionary algorithms such as genetic algorithm, genetic programming to address this issue both in time and frequency domain identification. The related literature can be found in~\cite{Billings:2013,Ljung:1999,Fonseca:1993,Swain:Billings:1998,Mendez:2001,Rodriguez:Fonseca:2004,Kar:Swain:2009,Baldacchino:Visakan:2012} and the references there in.  

In this study, a novel 2-D particle swarm algorithm~\cite{Hafiz:Swain:2017a} has been applied for selecting the correct structure of the NARX model of the nonlinear systems. The proposed algorithm explicitly include the cardinality (\textit{i.e.}, the number of terms)  information in to the search process by extending the dimension of classical single dimensional particle swarm algorithm. This is different from the least squares based orthogonal search algorithms developed by Billings and co-workers where the structure selection is carried out using an error-reduction-ratio (ERR) test which is computed either from the one-step ahead prediction or simulated prediction of the output~\cite{Billings:2013,Baldacchino:Visakan:2012}. 

The rest of the article is organized as follows: NARX model of the non-linear systems and the 2D learning approach are briefly described in Section~\ref{s:NARX} and~\ref{s:2DL}. The framework of this study is described in Section~\ref{s:IF}. The results are discussed at length in Section~\ref{s:res}, followed by the conclusions in Section~\ref{s:con}.

\section{The Polynomial NARX Model}
\label{s:NARX}
A wide class of nonlinear system can be represented by a polynomial nonlinear auto-regressive moving average with exogenous (NARMAX) model given by,
\begin{align*}
    y(k) ={} & F^{N_l} \ [ \ y(k-1),\ldots,y(k-n_y),u(k-1),\ldots, \\
      {} & ~~~~~~~  u(k-n_u), e(k-1),\ldots,e(k-n_e) \ ]+e(k)
\end{align*}
where $y(k)$, $u(k)$ and $e(k)$ represent the output, input and noise respectively at time intervals $k$, $n_y$, $n_u$ and $n_e$ are corresponding lags and $F^{N_l}[.]$ is some nonlinear function of degree $N_l$. The polynomial NARX model is a subclass of NARMAX model where the noise terms are absent and is given by,
\begin{align*}
y(k) = {} & F^{N_l} \ [ \ y(k-1),\ldots,y(k-n_y),u(k-1),\ldots,\\ 
       {} & ~~~~~~~ u(k-n_u) \ ]+e(k)
\end{align*}
The \textit{total number of possible terms} or \textit{model size} ($N_t$) of the NARMAX model is given by
\begin{align}
\label{eq:Nt}
N_t & = \sum_{i=0}^{N_l} n_i, \ n_0=1 \\
n_i & = \frac{n_{i-1}(n_y+n_u+n_e+i-1)}{i}, \ i=1,\ldots, N_l
\end{align}
This model is essentially linear-in-parameters and can be expressed as
\begin{align*}
    y(k) & = \sum_{i=1}^{N_t} \theta_i x_i(k) + e(k)
\end{align*}
where , $x_1(k) = 1$
\begin{align*}
    x_i(k) & = & \prod_{j=1}^{p_y}y(k-n_{y_j})\prod_{k=1}^{q_u}u(k-n_{u_k}) \prod_{m=1}^{r_e}e(k-n_{e_m})
\end{align*}
where, $i=2,\ldots, N_t$; $p_y,q_u,r_e \geq 0$; $1\leq p_y+q_u+r_e \leq N_l$; $1 \leq n_{y_j}\leq n_y$;$1 \leq n_{u_k}\leq n_u$; $1 \leq n_{e_m}\leq n_e$; and $N_l$ is the degree of polynomial expansion. For the polynomial NARX model, which does not contain any noise terms, $r_e=0$ and this indicates that $x_i(k)$ contains no $e(.)$ terms. 

\section{Two-Dimensional (2D) Particle Swarms}
\label{s:2DL}

The Two-Dimensional (2D) learning framework was developed for the particle swarms to effectively address the feature selection problem~\cite{Hafiz:Swain:2017a}. The core idea of this approach is to integrate the information about the cardinality (\textit{i.e.}, the number of features) into the search process. This learning approach has been applied to address the power quality problems~\cite{Hafiz:Swain:2019} and has been shown to perform better than some of the existing approaches~\cite{Zhang:Swain:2007,Naik:Hafiz:2016,Hafiz:Swain:2017b}. In this study, it has been shown that this learning approach can easily be applied to address the model structure selection problem, as in essence, the main objective of both the feature selection and the structure detection are similar, \textit{i.e.}, \textit{detect a parsimonious model} or \textit{remove redundant features/terms}. 

The following subsections briefly describe the 2D learning approach. More details about this approach can be found in~\cite{Hafiz:Swain:2017a}.

\subsection{Learning Philosophy}

The selection of a parsimonious model in the problems such as feature selection or structure selection involves mainly two issues : 1) How many attributes (\textit{e.g.}, \textit{features} or \textit{terms}) are required? 2) Which attributes shall be selected? The main philosophy of the 2D learning approach is to explicitly use the information about the cardinality (\textit{i.e.} the number of attributes/features/terms) as an additional learning dimension in order to facilitate informed decision on both issues. 

To understand the 2D learning approach, consider a structure selection problem with a total of `$N_t$' possible terms. For this problem, a particle position (model), `$x$', is usually represented as a $N_t$-dimensional binary string wherein selected terms are represented by bit `$1$'. 

Further, in the conventional Binary PSO and its variants, the velocity of a particle is given by,
\begin{align*}
    v & =\begin{bmatrix} p_{1} & p_{2} & \dots & p_{N_t}  \end{bmatrix}
\end{align*}
where, $ \{ p_1, \dots p_{N_t} \} $ represents the selection likelihood of the corresponding attributes. In other words, in each iteration, a new position (model) is derived \textit{only on the basis of selection likelihoods of the attributes}. 

In contrast, in 2D learning, the selection likelihood of both cardinality and attributes are considered to derive a new position. This is achieved by extending the learning dimension and separately storing selection likelihood of cardinality and attributes in a two-dimensional matrix of size $(2\times N_t)$ as follows:
\begin{align*}
v & =\begin{bmatrix} p_{11} & p_{12} & \dots & p_{1N_t} \\                     
                     p_{21} & p_{22} & \dots & p_{2N_t} \end{bmatrix}, \ \text{where, \ } v \in \mathbb{R}_{2 \times N_t}
\end{align*}
The elements in the first row of $v$ store the selection likelihoods of cardinality. For example, `$p_{12}$' gives the probability of selecting $2$ attributes in a new model. Similarly, the elements in the second row give the selection likelihood of the corresponding attributes, \textit{e.g.}, the selection probability of the `$3^d$' attribute is given by `$p_{23}$'.  


\subsection{Velocity Update}

The 2D learning approach was developed as a generalized learning framework to adapt any continuous PSO variant (\textit{i.e.}, $x \in \mathbb{R}$) for the model selection task (\textit{i.e.}, $x \in \mathbb{N}$). However, the results of comparative analysis of various PSO variants~\cite{Hafiz:Abdennour:2016} indicate that Unified Particle Swarm Optimization (UPSO)~\cite{Parsopoulos:Konstantinos:2004} is more suitable for the problem in the discrete domain (\textit{i.e.}, $x \in \mathbb{N}$) and therefore it is selected in this study.

In essence, UPSO combines the \textit{`global'} and \textit{`local'} variants of the conventional PSO through the unification factor, `$u$'. Following the 2D learning approach, UPSO is adapted for the structure selection task as follows:
\begin{align}
\label{eq:UPSOorig} 
v_{i}  & = (u \times v_{gi})+((1-u) \times v_{li})\\
\label{eq:UPSOupdated1}
\text{where, \ } v_{gi} & = (\omega \times v_{i}) + (c_1r_1 \times L_{cog}) \nonumber \\ 
                        & \; \; \; + (c_2r_2 \times L_{soc,1}) + (\Delta_{i} \times L_{self}) \\ 
\label{eq:UPSOupdated2}
v_{li} & = (\omega \times v_{i}) + (c_1r_1 \times L_{cog}) \nonumber \\
          & \; \; \; + (c_2r_2 \times L_{soc,2}) + (\Delta_{i} \times L_{self})
\end{align}
where, `$\omega$' is \textit{inertia weight} and $[c_1, c_2]$ denotes \textit{acceleration constants}. 

Further, in the 2D learning, a \textit{learning set}, `$L$', is derived from each learning exemplar, \textit{e.g.}, \textit{personal best} ($pbest$), \textit{global best} ($gbest$), \textit{neighborhood best} ($nbest$). For example, $L_{cog}$, $L_{soc,1}$, $L_{soc,2}$ and $L_{self}$ are the learning sets derived from $pbest$, $gbest$, $nbest$ and the particle position $x_i$, respectively. Note that the \textit{ring topology} is being used to define the neighborhood.

The learning set is a two-dimensional binary matrix of size ($2\times N_t$). Similar to the velocity matrix, learning sets store the learning about cardinality and attribute in separate rows. More details about the learning process and the derivation of the learning sets can be found in~\cite{Hafiz:Swain:2017a}. 

Note that, in addition to the learning exemplars, the particle position ($x$) is used to extract a learning set, referred as a \textit{self learning set}. The influence of the self learning set is controlled by the `$\Delta$' which is given by,
\begin{gather}
\label{eq:fitfeedback}
       \Delta_i= \begin{cases}
                + \delta_i, &   if \; \; \frac{f_{i}^{t}}{f_{i}^{t-1}}<1\\
                - \delta_i, &   otherwise
                 \end{cases}\\
\label{eq:delta}
\text{where,\ }      \delta_i = 1- \frac{f_{i}^{t}}{max(F^{t})}\nonumber
\end{gather}
where, `$f_{i}^{t}$' and `$F^t$' are respectively the fitness of the $i^{th}$ particle and the vector containing the fitness of the entire swarm at iteration $t$. The objective here is to adjust the selection likelihood of the terms included in the particle position as per : 1) the relative improvement over other particles  and 2) the relative improvement in the self-fitness over the previous iteration.  

\begin{algorithm}[!t]
    \small
    \SetKwInOut{Input}{Input}
    \SetKwInOut{Output}{Output}
    \SetKwComment{Comment}{*/ \ \ \ }{}
    \Input{$v_i$}
    \Output{$x_i$}
    \BlankLine
    Set the new position $x_i$ to an $n$-dimensional null vector, \textit{i.e.}, $x_i=\{ 0 \dots 0 \}$ \\
    \BlankLine
    Isolate the selection likelihood of the \emph{cardinality} and \emph{feature} into respective vectors, `$\rho$' and `$\sigma$' using (\ref{eq:sig_rho})
    \BlankLine
    \BlankLine
    \Comment*[h] {roulette wheel selection of the cardinality, ($\xi_i$)}\\
    Evaluate accumulative probabilities, $\rho_{\Sigma,j} = \sum \limits_{k=1}^{j} \rho_{k}, \ \  j = 1\dots n$. \nllabel{line:pos1}\\
    \BlankLine
    Generate a random number, $r \in [0,\rho_{\Sigma,n}]$. \nllabel{line:pos2}\\
    \BlankLine
    Determine `$j$' such that $\rho_{\Sigma,j-1}<r<\rho_{\Sigma,j}$, this gives the size of the subset $\xi_i$, \textit{i.e.}, $\xi_i=j$. \nllabel{line:pos3}\\
    \BlankLine
    \BlankLine
    \Comment*[h]{Selection of the terms}\\
    Rank the terms on the basis of their \emph{likelihood} `$\sigma_j$' and store the term rankings in vector `$\tau$' \nllabel{line:fs1}\\
    \BlankLine
    \For{j = 1 to n} 
        { \If{$\tau_{j} \leq \xi_i$}
            {$x_{i,j}=1$}
        } \nllabel{line:fs2}
\caption{2-D learning approach to the position update of the $i^{th}$ particle}
\label{fig:posprop}
\end{algorithm}
\begin{algorithm}[!t]
    \small
    \SetKwInOut{Input}{Input}
    \SetKwInOut{Output}{Output}
    \SetKwComment{Comment}{*/ \ \ \ }{}
    \Input{\textit{System input-output measurement data}}
    \Output{\textit{Model Structure}}
    \BlankLine
    Set the search parameters: $c_1, \ c_2, \  \omega, \ u_0, \ u_f \ \& \  RG$ \\
    \BlankLine
    Randomly initialize the swarm of `$ps$' number of particles, $X=\{ x_1 \dots x_{ps} \}$ \\
    Initialize the velocity ($(2 \times n)$ matrix) of each particle by uniformly distributed random numbers in [0,1] \\
    \BlankLine
    Evaluate the fitness of the swarm, $pbest$ and $gbest$ \\
    \BlankLine
    \For{t = 1 to iterations}
        { 
            Evaluate unification factor, $u(t) = u_0 + \frac{(u_f - u_0) \times t}{Max \ Iteration}$ \\
            \BlankLine
            \Comment*[h]{Swarm Update}\\
            \BlankLine
            \For{i = 1 to ps} 
            {
                \BlankLine
                \Comment*[h]{Stagnation Check}\\
                \BlankLine
                \If{$ count_i \geq RG$}
                    {Re-initialize the velocity of the particle\\
                    Set $count_i$ to zero}
                \BlankLine
                Update the velocity of the $i^{th}$ particle as per~(\ref{eq:UPSOorig}), (\ref{eq:UPSOupdated1}) and~(\ref{eq:UPSOupdated2}) \\
                Update the position of the $i^{th}$ particle following Algorithm-\ref{fig:posprop}
            }
            \BlankLine
            Store the old fitness of the swarm in `$F$'\\
            \BlankLine
            Evaluate the swarm fitness\\
            \BlankLine
            Update personnel and global best position, $pbest$, $gbest$\\             
            \BlankLine
            \Comment*[h]{Stagnation Check}\\ 
            \BlankLine
            \For{i = 1 to ps}
                { \If{$pbestval_i^{ \ t} \geq pbestval_i^{ \ t-1}$}
                        {$count_i=count_i+1$}
                }
        
        }
\caption{Pseudo code of 2D-UPSO algorithm for the model structure selection}
\label{fig:2D-UPSO}
\end{algorithm}
\begin{table*}[!t]
  \centering
  \caption{Test Non-linear Systems}
  \label{t:sys}%
  \begin{adjustbox}{max width=0.95\textwidth}
  \scriptsize
 \begin{threeparttable}
    \begin{tabular}{ccc}
    \toprule
    \multirow{2}{*}{\textbf{System}} & \multirow{2}{*}{\textbf{Known Structure}} & \textbf{NARX Parameters$^\dagger$} \\
\cmidrule{3-3}          &       & $\boldmath[n_u,n_y,N_l,N_t]$ \\
    \midrule
    $S1$    &  $y(k) = 0.5y(k-1) + 0.3u(k-1) + 0.3y(k-1)u(k-1) + 0.5u(k-1)^2$     &  $[5,5,2,66]$\\ [1.2ex]
    $S2$    &  $y(k) = 0.5 + 0.5y(k-1) + 0.8u(k-2) + u(k-1)^2 - 0.05y(k-2)^2$     &  $[5,5,2,66]$\\[1.2ex]
    $S3$    &  $y(k) = 0.8y(k-1) + 0.4u(k-1) + 0.4u(k-1)^2 + 0.4u(k-1)^3$         &  $[3,3,3,84]$\\[2ex]
    
    $S4$    &  \makecell{$y(k) = 0.8833u(k-1) + 0.0393u(k-2) + 0.8546u(k-3) + 0.8528u(k-1)^2 +       0.7582u(k-1)u(k-2)$ \\[0.7ex] 
                         $ + 0.1750u(k-1)u(k-3) + 0.0864u(k-2)^2 + 0.4916u(k-2)u(k-3)  + 0.0711u(k-3)^2 $ \\[0.7ex]
                         $ - 0.0375y(k-1) - 0.0598y(k-2) - 0.0370y(k-3) - 0.0468y(k-4) - 0.0476y(k-1)^2 - 0.0781y(k-1)y(k-2)$\\[0.7ex]
                         $ - 0.0189y(k-1)y(k-3) - 0.0626y(k-1)y(k-4) - 0.0221y(k-2)^2 - 0.0617y(k-2)y(k-3)$\\[0.7ex]
                         $ - 0.0378y(k-2)y(k-4) - 0.0041y(k-3)^2 - 0.0543y(k-3)y(k-4) - 0.0603y(k-4)^2$ }       &  $[5,5,2,66]$\\
    
    \bottomrule
    \end{tabular}%
    \begin{tablenotes}
      \scriptsize
      \item $\dagger$ `$n_u$', `$n_y$', `$N_l$' and `$N_t$' respectively denotes \textit{input lag}, \textit{output lag}, \textit{degree of non-linearity} and \textit{total number of terms} of NARX model 
    \end{tablenotes}
  \end{threeparttable}
 \end{adjustbox}
\end{table*}%
\subsection{Position Update}

To understand the position update process, consider the velocity of an $i^{th}$ particle for a structure selection problem with $N_t=5$ as follows:
\begin{align}
\label{eq:example}
v_i & =\begin{bmatrix} p_{11}^i & p_{12}^i & p_{13}^i & p_{14}^i & p_{15}^i \\                     
                       p_{21}^i & p_{22}^i & p_{23}^i & p_{24}^i & p_{25}^i \end{bmatrix}\\
 & = \ \begin{bmatrix} 0.82 & 2.53 & 2.22 & 0.28 & 0.95 \\
                       1.61 & 1.88 & 0.80 & 1.33 & 2.88 \end{bmatrix}
\end{align}
To integrate the information about both the cardinality and the attributes/terms, the position update is carried out in two stages. In the first stage, the cardinality of the new position is determined. Consequently, the beneficial terms are selected as per the selected cardinality. This procedure is explained in the following example. 

For the sake of clarity, the velocity of the $i^{th}$ particle in (\ref{eq:example}) is segregated into two vectors $\rho$ and $\sigma$ as follows:
\begin{gather}
\label{eq:sig_rho}
v_i = \begin{bmatrix} \rho & \sigma \end{bmatrix} ^ T\\ 
\text{where, \ } \rho = \begin{bmatrix} p_{11}^i & \dots  & p_{15}^i \end{bmatrix} \ \text{and \ } \sigma = \begin{bmatrix} p_{21}^i & \dots  & p_{25}^i \end{bmatrix}\nonumber\\
\text{ which gives, \ }  \rho = \begin{bmatrix} 0.82 & 2.53 & 2.22 & 0.28 & 0.95 \end{bmatrix}\nonumber \\ 
\text{and \ }  \sigma = \begin{bmatrix} 1.61 & 1.88 & 0.80 & 1.33 & 2.88 \end{bmatrix}\nonumber
\end{gather}

The cardinality of the new position is selected following the roulette wheel approach as outlined in Lines~\ref{line:pos1}-\ref{line:pos3}, Algorithm~\ref{fig:posprop}. For the $i^{th}$ particle considered in (\ref{eq:example}) the cardinality of the new position, $\xi_i$ is derived as follows: 
\begin{enumerate}
    \item Accumulative likelihoods, \\
    $\rho_\Sigma = \begin{bmatrix} 0.82 & 3.35 & 5.57 & 5.85 & 6.80 \end{bmatrix}$
    \item Let random number $r\in[0,\rho_{\Sigma,n}]$ be $3.5$, which gives $\xi_i=3$ as $\rho_{\Sigma,2}<r<\rho_{\Sigma,3}$.
\end{enumerate}

The next step is to select $3$ most beneficial terms (as $\xi_i=3$). For this purpose, the steps outlined in Line~\ref{line:fs1}-\ref{line:fs2}, Algorithm~\ref{fig:posprop} are followed.
\begin{enumerate}
    \item Ranking of the terms,\\
            $\sigma = \begin{bmatrix} 1.61 & 1.88 & 0.80 & 1.33 & 2.88 \end{bmatrix}$ \\
            which gives, $\tau = \begin{bmatrix} 3 & 2 & 5 & 4 & 1 \end{bmatrix}$
    \item Selection of terms as per $\xi_i$,\\
            $\xi_i=3$ and $\tau = \begin{bmatrix} 3 & 2 & 5 & 4 & 1 \end{bmatrix}$ \\
            which gives, $x_i= \begin{bmatrix} 1 & 1 & 0 & 0 & 1 \end{bmatrix}$
\end{enumerate}

\section{Investigation Framework}
\label{s:IF}

The main objective of this study is to evaluate the search performance of 2D particle swarms on the structure selection problem. For this purpose, the performance of the 2D-UPSO has been compared with Binary Genetic Algorithm (GA) over several known non-linear systems, as shown in Table~\ref{t:sys}. 

The other objective is to evaluate the effects of the measurement noise and the choice of the fitness function on the search performance. For this purpose, the search performance is quantified using two performance metrics and compared over two distinct fitness functions at $3$ noise levels, SNR$=[50dB, 40dB, 30dB]$.

The following subsections provide the details about compared algorithms, fitness function and the performance metrics being used in this investigation.

\subsection{Search Setup}

To evaluate the efficacy of 2D particle swarms, the performance of 2D-UPSO has been compared with GA. Both algorithms were implemented in MATLAB. Due to stochastic nature of these algorithms, $40$ independent runs were carried out for each system. Each run was set to terminate after $6000$ Function Evaluations (FEs). The parameter settings of the compared algorithms are shown in Table~\ref{t:sparam}.

\begin{table}[!b]
\centering
\caption{Search Parameter Settings}
\label{t:sparam}
\begin{adjustbox}{max width=0.8\textwidth}
\scriptsize
\begin{threeparttable}
\begin{tabular}{c c} 
 \toprule
  \textbf{Algorithm} &  \textbf{Search Parameters}  \\ [1ex]
 \midrule
GA \cite{Siedlecki:Sklansky:1989,Grefenstette:1986} &  $N=80$, $p_c=0.45$, $p_m=0.01$\\ [1.5ex]
2D-UPSO\cite{Hafiz:Swain:2017a}                     &  \makecell{$ps=30$, $\omega=0.729$, $[c_1,c_2]=[1.49,1.49]$ \\ $u:0.2 \rightarrow 0.7$, $RG=10$} \\ [0.8ex]
\bottomrule
\end{tabular}
\begin{tablenotes}
      \scriptsize
      \item `$N$' - GA population size, $p_c,p_m$ - crossover and mutation probability
      \item `$ps$' - swarm size, `$\omega$' - inertia weight, `$c_1,c_2$' - acceleration constants 
      \item `$u$' - unification factor, `$RG$' - refresh gap
    \end{tablenotes}
  \end{threeparttable}
 \end{adjustbox}
\end{table}
\begin{table*}[!t]
\begin{minipage}[c]{.5\linewidth}
  \centering  
  \scriptsize 
  \caption{Effect of the fitness function - $S1$ $^\dagger$} 
  \label{t:fit1}%
   \begin{adjustbox}{max width=0.88\textwidth}
    \begin{threeparttable}
    
    \begin{tabular}{cccccccc}
    \toprule
    \multirow{2}{*}{\makecell{\textbf{Fitness Function,} \\ \boldmath$J(\cdotp)$}} & \multirow{2}{*}{\textbf{SNR}} & \multicolumn{4}{c}{\textbf{\boldmath$\nu$ of System Terms $^\ddagger$}} & \multicolumn{2}{c}{\textbf{Spurious Terms}} \\
    \cmidrule{3-8} &  & \boldmath$T_1$ & \boldmath$T_2$ & \boldmath$T_3$ & \boldmath$T_4$ & \boldmath$r$ & \boldmath$\nu_{max}$ \\
    \midrule
    \multirow{3}{*}{BIC} & 50 dB & 1 & 1 & 1 & 1 & 0.15  & 0.28 \\
          & 40 dB & 1 & 1 & 1 & 1 & 0.21  & 0.28 \\
          & 30 dB & 1 & 1 & 1 & 1 & 0.20  & 0.15 \\
    \midrule
    \multirow{3}{*}{\makecell{AIC \\ ($\varrho=2$)}} & 50 dB & 1 & 1 & 1 & 1 & 0.33  & 0.53 \\
          & 40 dB & 0.98  & 1 & 1 & 1 & 0.44  & 0.45 \\
          & 30 dB & 0.98  & 1 & 1 & 1 & 0.52  & 0.50 \\
    \midrule
    \multirow{3}{*}{\makecell{AIC \\ ($\varrho=2^3$)}} & 50 dB & 1 & 1 & 1 & 1 & 0.14  & 0.20 \\
          & 40 dB & 1 & 1 & 1 & 1 & 0.17  & 0.23 \\
          & 30 dB & 1 & 1 & 1 & 1 & 0.24  & 0.30 \\
    \midrule
    \multirow{3}{*}{\makecell{AIC \\ ($\varrho=2^6$)}} & 50 dB & 1 & 1 & 1 & 1 & 0.15  & 0.30 \\
          & 40 dB & 1 & 1 & 1 & 1 & 0.12  & 0.28 \\
          & 30 dB & 0.98  & 1 & 1 & 1 & 0.24  & 0.25 \\
    \midrule
    \multirow{3}{*}{\makecell{AIC \\ ($\varrho=2^8$)}} & 50 dB & 1 & 1 & 1 & 1 & 0.06 & 0.10 \\
          & 40 dB & 1 & 1 & 1 & 1 & 0.06  & 0.10 \\
          & 30 dB & 1 & 1 & 1 & 1 & 0.08  & 0.10 \\
    \bottomrule
    \end{tabular}%

    \begin{tablenotes}
    \scriptsize
    \item $\dagger$ `$r$' denotes the ratio of number of redundant terms to the total number of terms
    \smallskip
    \item $^\ddagger$ $ \{ T_1, \dots T_4 \} \in T_{system}$ denote the system terms and are given in Appendix~\ref{s:app} 
    \end{tablenotes}
    \end{threeparttable}
 \end{adjustbox} 
 \end{minipage}
 \hfill
  \begin{minipage}[c]{.5\linewidth}
  \centering 
  \scriptsize
  \caption{Effect of the fitness function - $S2$ $^\dagger$} 
  \label{t:fit2}%
   \begin{adjustbox}{max width=0.98\textwidth}
    \begin{threeparttable}
    
    \begin{tabular}{ccccccccc}
    \toprule
    \multirow{2}{*}{\makecell{\textbf{Fitness Function,} \\ \boldmath$J(\cdotp)$}} & \multirow{2}{*}{\textbf{SNR}} & \multicolumn{5}{c}{\textbf{\boldmath$\nu$ of System Terms (\boldmath$T_{system}$)$^\ddagger$}} & \multicolumn{2}{c}{\textbf{Spurious Terms}} \\
    \cmidrule{3-9} &  & \boldmath$T_1$ & \boldmath$T_2$ & \boldmath$T_3$ & \boldmath$T_4$ & \boldmath$T_5$ & \boldmath$r$ & \boldmath$\nu_{max}$ \\
    \midrule
    \multirow{3}{*}{BIC} & 50 dB & 1     & 0.98  & 1 & 1 & 1 & 0.38 & 0.45 \\
          & 40 dB & 1 & 0.98 & 1 & 1 & 1 & 0.36 & 0.38 \\
          & 30 dB & 1 & 1 & 1 & 1 & 1 & 0.48 & 0.50 \\
    \midrule
    \multirow{3}{*}{\makecell{AIC \\ ($\varrho=2$)}} & 50 dB & 1 & 0.88  & 1 & 1 & 1 & 0.53 & 0.83 \\
          & 40 dB & 1 & 0.90  & 1 & 1 & 1 & 0.61  & 0.50 \\
          & 30 dB & 1 & 0.78  & 1 & 1 & 1 & 0.68  & 0.65 \\
    \midrule
    \multirow{3}{*}{\makecell{AIC \\ ($\varrho=2^3$)}} & 50 dB & 1 & 0.98 & 1 & 1 & 1 & 0.39 & 0.38 \\
          & 40 dB & 1     & 0.95  & 1     & 1     & 1     & 0.47  & 0.33 \\
          & 30 dB & 1     & 1     & 1     & 1     & 1     & 0.41  & 0.58 \\
    \midrule
    \multirow{3}{*}{\makecell{AIC \\ ($\varrho=2^6$)}} & 50 dB & 1 & 0.93 & 1 & 1 & 1 & 0.38 & 0.33 \\
          & 40 dB & 1 & 1 & 1 & 1 & 1 & 0.36  & 0.40 \\
          & 30 dB & 1 & 0.95 & 1 & 1 & 1 & 0.30 & 0.38 \\
    \midrule
    \multirow{3}{*}{\makecell{AIC \\ ($\varrho=2^8$)}} & 50 dB & 1 & 1 & 1 & 1 & 1 & 0.08  & 0.10 \\
          & 40 dB & 1 & 1 & 1 & 1 & 1 & 0.09  & 0.08 \\
          & 30 dB & 0.98  & 0.95  & 1 & 0.90  & 1 & 0.20  & 0.10 \\
    \bottomrule
    \end{tabular}%

    \begin{tablenotes}
    \scriptsize
    \item $^{\dagger}$ - `$r$' denotes the ratio of number of redundant terms to the total number of terms
    \smallskip
    \item $^\ddagger$ $ \{ T_1, \dots T_5 \} \in T_{system}$ denote the system term and are given in Appendix~\ref{s:app}
    \end{tablenotes}
    \end{threeparttable}
 \end{adjustbox}   
 \end{minipage}
\end{table*}%
\subsection{Fitness Function}
The fitness function is the only link between the search algorithm and the problem being solved. Hence, the choice of fitness function is likely to have a significant impact on the search performance. In this study, to evaluate the `\textit{fitness}' of a candidate model (particle position) `$x$', \textit{sum-squared-error} (SSE) with respect to `\textit{model-predict output}' over validation data is evaluated,
\begin{align}
\label{eq:sse}
    e & = \sum \limits_{k=1}^{L} [ y_k - \hat{y_k} ]^2
\end{align}
where, `$L$' denotes the total number of validation data samples and `$\hat{y}$' denotes `\textit{model-predicted output}' corresponding to $x$. For each system shown in Table~\ref{t:sys}, 2000 pairs of input-output data, $(u,y)$, were generated. Out of these data, $30\%$ data were used for the validation purposes (\textit{i.e.}, $L=600$).

Note that minimization of `$e$' does not guarantee a parsimonious model. It is therefore necessary to assign a penalty for a higher cardinality (number of terms) to ensure the removal of the redundant terms. Hence, the structure selection is essentially multi-objective problem. To balance both search objectives (\textit{i.e.}, minimizing $e$ and cardinality) several information theoretic criteria have been proposed over the years such as Akaike Information Criterion (AIC), Minimum Description Length (MDL), Bayesian Information Criterion (BIC) and others~\cite{Billings:2013}. In this study, BIC (\ref{eq:bic}) and AIC (\ref{eq:aic}) are selected as the fitness function, $J(\cdotp)$,
\begin{align}
\label{eq:bic}
J(x) & = L \ln(e) + \ln(L) \xi \\
\label{eq:aic}
J(x) & = L \ln(e) + \varrho \xi
\end{align}
Here, `$x$' denotes a model structure (particle position) under consideration, `$\xi$' gives number of terms included in $x$ (or \textit{cardinality} of $x$). Further, $\varrho$ in AIC (\ref{eq:aic}) is a constant and requires a careful tuning. In this study, four distinct values of $\varrho$ have been used, $\varrho=[2, 2^3, 2^6, 2^8]$.

\subsection{Performance Metrics}
\label{s:pm}
The stochastic nature of the compared algorithm requires a multiple independent runs ($40$ in this study) for each system. Since, it is likely that each run provides a different model structure, a usual approach is to select the model structure with the best fitness out of all runs. In contrast, in this study a comprehensive approach is followed. For a given system, a model structure (selected terms) and its cardinality are stored at the end of each run of the algorithm. The objective is to evaluate the frequency of selection of each term (`$\nu$') over $40$ runs. The final model structure is obtained by including terms with $\nu$ over certain fixed threshold. The details are discussed in the following.

Let us define the following sets to quantify the search performance,
\begin{itemize}
    \item $T_{model}$ : set of all possible terms present in the search domain, \textit{i.e.}, it contains the complete possible model terms.
    \item $T_{system}$ : set of the terms which are present in the actual system, $T_{system} \subset T_{model}$
    \item $T_{algorithm}$ : set of terms selected by a search algorithm, $T_{algorithm} \subset T_{model}$
    \item $T_{spur}$ : set of \textit{spurious} terms which are selected by the search algorithm but which are not present in the actual system, \textit{i.e.}, $T_{spur} = T_{algorithm} - T_{system}$ 
\end{itemize}

In order to quantify the search performance, selection frequency, $\nu$ (\ref{eq:nu}), of system ($T_{system}$) and spurious ($T_{spur}$) terms is observed over 40 runs. It is desirable to have a higher selection frequency of the system terms and fewer spurious terms.
\begin{equation}
\label{eq:nu}
\nu_i = \frac{\textit{selection frequency of the \ } i^{th}\textit{term}}{\textit{total number of runs}}
\end{equation}
To extract the final structure, a threshold can be fixed at any suitable value in $[0,1]$, \textit{e.g.} with threshold at $0.9$, only the terms $\nu \geq 0.9$ are included in the final structure.

\begin{table}[!b]
  \centering 
  \scriptsize
  \caption{Effect of the fitness function - $S3$ $^\dagger$} 
  \label{t:fit3}%
   \begin{adjustbox}{max width=0.47\textwidth}
    \begin{threeparttable}
    
    \begin{tabular}{cccccccc}
    \toprule
    \multirow{2}{*}{\makecell{\textbf{Fitness} \\[0.5ex] \textbf{Function}, \boldmath$J(\cdotp)$}} & \multirow{2}{*}{\textbf{SNR}} & \multicolumn{4}{c}{\textbf{\boldmath$\nu$ of System Terms (\boldmath$T_{system}$)$^\ddagger$}} & \multicolumn{2}{c}{\textbf{Spurious Terms}} \\

    \cmidrule{3-8} &  & \boldmath$T_1$ & \boldmath$T_2$ & \boldmath$T_3$ & \boldmath$T_4$ & \boldmath$r$ & \boldmath$\nu_{max}$ \\
    \midrule
    \multirow{3}{*}{BIC} & 50 dB & 1     & 1     & 1     & 1     & 0.24  & 0.38 \\
          & 40 dB & 1     & 1     & 1     & 1     & 0.26  & 0.40 \\
          & 30 dB & 0.95  & 1     & 1     & 1     & 0.48  & 0.50 \\
    \midrule
    \multirow{3}{*}{\makecell{AIC \\ ($\varrho=2$)}} & 50 dB & 1     & 1     & 1     & 1     & 0.64  & 0.90 \\
          & 40 dB & 0.95  & 1     & 1     & 1     & 0.62  & 0.70 \\
          & 30 dB & 0.95  & 1     & 1     & 1     & 0.85  & 0.78 \\
    \midrule
    \multirow{3}{*}{\makecell{AIC \\ ($\varrho=2^3$)}} & 50 dB & 1     & 1     & 1     & 1     & 0.24  & 0.23 \\
          & 40 dB & 1     & 1     & 1     & 1     & 0.19  & 0.30 \\
          & 30 dB & 0.98  & 1     & 1     & 1     & 0.31  & 0.45 \\
    \midrule
    \multirow{3}{*}{\makecell{AIC \\ ($\varrho=2^6$)}} & 50 dB & 0.98  & 1     & 1     & 1     & 0.18  & 0.20 \\
          & 40 dB & 1     & 1     & 1     & 1     & 0.15  & 0.15 \\
          & 30 dB & 1     & 1     & 1     & 1     & 0.07  & 0.13 \\
    \midrule
    \multirow{3}{*}{\makecell{AIC \\ ($\varrho=2^8$)}} & 50 dB & 1     & 1     & 1     & 1     & 0.05  & 0.13 \\
          & 40 dB & 1     & 1     & 1     & 1     & 0.05  & 0.08 \\
          & 30 dB & 1     & 1     & 1     & 1     & 0.07  & 0.13 \\
    \bottomrule
    \end{tabular}%

    \begin{tablenotes}
    \scriptsize
    \item $^{\dagger}$ - `$r$' denotes the ratio of number of redundant terms to the total number of terms
    \smallskip
    \item $^\ddagger$ $ \{ T_1, \dots T_5 \} \in T_{system}$ denote the system term and are given in Appendix~\ref{s:app}
    \end{tablenotes}
    \end{threeparttable}
 \end{adjustbox}    
\end{table}%
\begin{table*}[!t]
  \centering 
  \small
  \caption{Effect of the fitness function - $S4$ $^\dagger$} 
  \label{t:fit4}%
   \begin{adjustbox}{max width=0.999\textwidth}
    \begin{threeparttable}
    
    \begin{tabular}{ccccccccccccccccccccccccccc}
    \toprule
     \multirow{2}{*}{\makecell{\textbf{Fitness} \\ \textbf{Function,} \\[0.75ex] \boldmath$J(\cdotp)$}} & \multirow{2}{*}{\textbf{SNR}} & \multicolumn{23}{c}{\textbf{Selection Frequency (\boldmath$\nu$) of System Terms (\boldmath$T_{system}$)$^\ddagger$}} & \multicolumn{2}{c}{\textbf{Spurious Terms}} \\[1.2ex]
    \cmidrule{3-27} & & \boldmath$T_{1}$ & \boldmath$T_{2}$ & \boldmath$T_{3}$ & \boldmath$T_{4}$ & \boldmath$T_{5}$ & \boldmath$T_{6}$ & \boldmath$T_{7}$ & \boldmath$T_{8}$ & \boldmath$T_{9}$ & \boldmath$T_{10}$ & \boldmath$T_{11}$ & \boldmath$T_{12}$ & \boldmath$T_{13}$ & \boldmath$T_{14}$ & \boldmath$T_{15}$ & \boldmath$T_{16}$ & \boldmath$T_{17}$ & \boldmath$T_{18}$ & \boldmath$T_{19}$  & \boldmath$T_{20}$ & \boldmath$T_{21}$ & \boldmath$T_{22}$ & \boldmath$T_{23}$ & \boldmath$r$ & \boldmath$\nu_{max}$ \\ [1.2ex]
    \midrule
    \multirow{3}{*}{BIC} & $50dB$ & 1     & 1     & 1     & 1     & 1     & 1     & 1     & 1     & 1     & 1     & 1     & 1     & 1     & 1     & 1     & 1     & 1     & 1     & 1     & 1     & 1     & 1     & 1     & 0.64  & 0.88 \\
          & $40dB$ & 1     & 1     & 1     & 1     & 1     & 1     & 1     & 1     & 1     & 1     & 1     & 1     & 1     & 1     & 1     & 1     & 1     & 1     & 1     & 1     & 1     & 1     & 1     & 0.65  & 0.95 \\
          & $30dB$ & 0.98  & 1     & 1     & 1     & 1     & 1     & 1     & 1     & 1     & 1     & 1     & 1     & 1     & 1     & 0.98  & 1     & 1     & 1     & 1     & 1     & 1     & 1     & 1     & 0.62  & 0.90 \\[0.8ex]
    \midrule
    \multirow{3}{*}{\makecell{AIC \\ ($\varrho=2$)}} & $50dB$ & 1     & 1     & 1     & 1     & 1     & 1     & 1     & 1     & 1     & 1     & 1     & 1     & 1     & 1     & 1     & 1     & 1     & 1     & 1     & 1     & 1     & 1     & 1     & 0.64  & 0.98 \\
          & $40dB$ & 1     & 1     & 1     & 1     & 1     & 1     & 1     & 1     & 1     & 1     & 1     & 1     & 1     & 1     & 1     & 1     & 1     & 1     & 1     & 1     & 1     & 1     & 1     & 0.64  & 0.98 \\
          & $30dB$ & 0.93  & 1     & 1     & 1     & 1     & 0.98  & 1     & 1     & 1     & 1     & 1     & 1     & 1     & 1     & 1     & 1     & 1     & 1     & 1     & 1     & 0.98  & 1     & 1     & 0.64  & 0.98 \\[0.8ex]
    \midrule
    \multirow{3}{*}{\makecell{AIC \\ ($\varrho=2^3$)}} & $50dB$ & 1     & 1     & 1     & 1     & 1     & 1     & 1     & 1     & 1     & 1     & 1     & 1     & 1     & 1     & 1     & 1     & 1     & 1     & 1     & 1     & 1     & 1     & 1     & 0.64  & 0.93 \\
          & $40dB$ & 1     & 1     & 1     & 1     & 1     & 1     & 1     & 1     & 1     & 1     & 1     & 1     & 1     & 1     & 1     & 1     & 1     & 1     & 1     & 1     & 1     & 1     & 1     & 0.62  & 0.93 \\
          & $30dB$ & 0.95  & 1     & 1     & 1     & 1     & 1     & 1     & 1     & 1     & 1     & 1     & 1     & 1     & 1     & 1     & 1     & 1     & 1     & 1     & 1     & 0.98  & 1     & 1     & 0.64  & 0.93 \\[0.8ex]
    \midrule
    \multirow{3}{*}{\makecell{AIC \\ ($\varrho=2^6$)}} & $50dB$ & 1     & 1     & 1     & 1     & 1     & 1     & 1     & 1     & 1     & 1     & 1     & 1     & 1     & 1     & 1     & 1     & 1     & 1     & 1     & 1     & 1     & 1     & 1     & 0.64  & 0.90 \\
          & $40dB$ & 1     & 1     & 1     & 1     & 1     & 1     & 1     & 1     & 1     & 1     & 1     & 1     & 1     & 1     & 1     & 1     & 1     & 1     & 1     & 1     & 1     & 1     & 1     & 0.64  & 0.90 \\
          & $30dB$ & 0.85  & 0.98  & 0.98  & 1     & 1     & 0.95  & 1     & 1     & 1     & 1     & 1     & 1     & 1     & 1     & 0.85  & 1     & 1     & 1     & 1     & 1     & 0.95  & 1     & 0.9   & 0.62  & 0.80 \\[0.8ex]
    \midrule
    \multirow{3}{*}{\makecell{AIC \\ ($\varrho=2^8$)}} & $50dB$ & 0.05  & 0.63  & 0     & 0.33  & 1     & 0     & 0.20  & 0     & 0     & 0     & 0     & 0     & 0     & 0     & 0     & 0     & 0     & 0.35  & 0.35  & 0     & 0     & 0     & 0     & 0     & 0 \\
          & $40dB$ & 0.08  & 0.48  & 0.10  & 0.35  & 1     & 0     & 0.28  & 0     & 0     & 0     & 0     & 0     & 0     & 0     & 0     & 0     & 0     & 0.43  & 0.43  & 0     & 0     & 0     & 0     & 0     & 0 \\
          & $30dB$ & 0.20  & 0.38  & 0.05  & 0.38  & 1     & 0     & 0.15  & 0     & 0     & 0     & 0     & 0     & 0     & 0     & 0     & 0     & 0     & 0.60  & 0.43  & 0     & 0     & 0     & 0     & 0     & 0 \\
    \bottomrule
    \end{tabular}%

    \begin{tablenotes}
    \small
    \item $^{\dagger}$ `$r$' denotes the ratio of number of redundant terms to the total number of terms
    \smallskip
    \item $^\ddagger$ $ \{ T_1, \dots T_{23} \} \in T_{system}$ denote the system terms and are given in Appendix~\ref{s:app}
    \end{tablenotes}
    \end{threeparttable}
 \end{adjustbox}    
\end{table*}%
Further, to account the spurious terms selected by the algorithm, the following two metrics are used:
\begin{align}
\label{eq:num}
\nu_{max} & = \max\limits_{j} \ \nu_j, \text{ \ $j^{th}$ term \ } \in T_{spur} \\ 
\label{eq:r}
r & =\frac{n_{spur}}{N_t}
\end{align}
where, `$n_{spur}$' denotes the number of spurious terms identified by the search algorithm; `$N_t$' denotes the model size which is given by~(\ref{eq:Nt}). The metric $\nu_{max}$ gives the maximum selection frequency of the spurious terms. A lower value $\nu_{max}$ is desirable, as it ensures that $\nu$ of the spurious terms will be less than selection threshold. Similarly, a lower value of metric `$r$' is desirable, as it indicates that a small number of spurious terms are selected by the search algorithm. 

\section{Results}
\label{s:res}

The main objective of this study is to investigate the search performance of 2D particle swarms on the structure selection problem. Further, we are also interested in evaluating the effects of measurement noise and the choice of the fitness function on the search performance. For this purpose, two well-known information theoretic criteria such as AIC and BIC have been used as a fitness function. In the first stage of the investigation, we compare these criteria for structure selection problem in the presence of various levels of measurement noise. The results indicate that BIC is more suitable for this task (as discussed in Section~\ref{s:r1}). Consequently, the search performance of 2D-UPSO and GA is compared with BIC as the fitness function during the second stage of the investigation. 
\begin{figure*}[!t]
\centering
\begin{subfigure}{.37\textwidth}
  \centering
  \includegraphics[width=\textwidth]{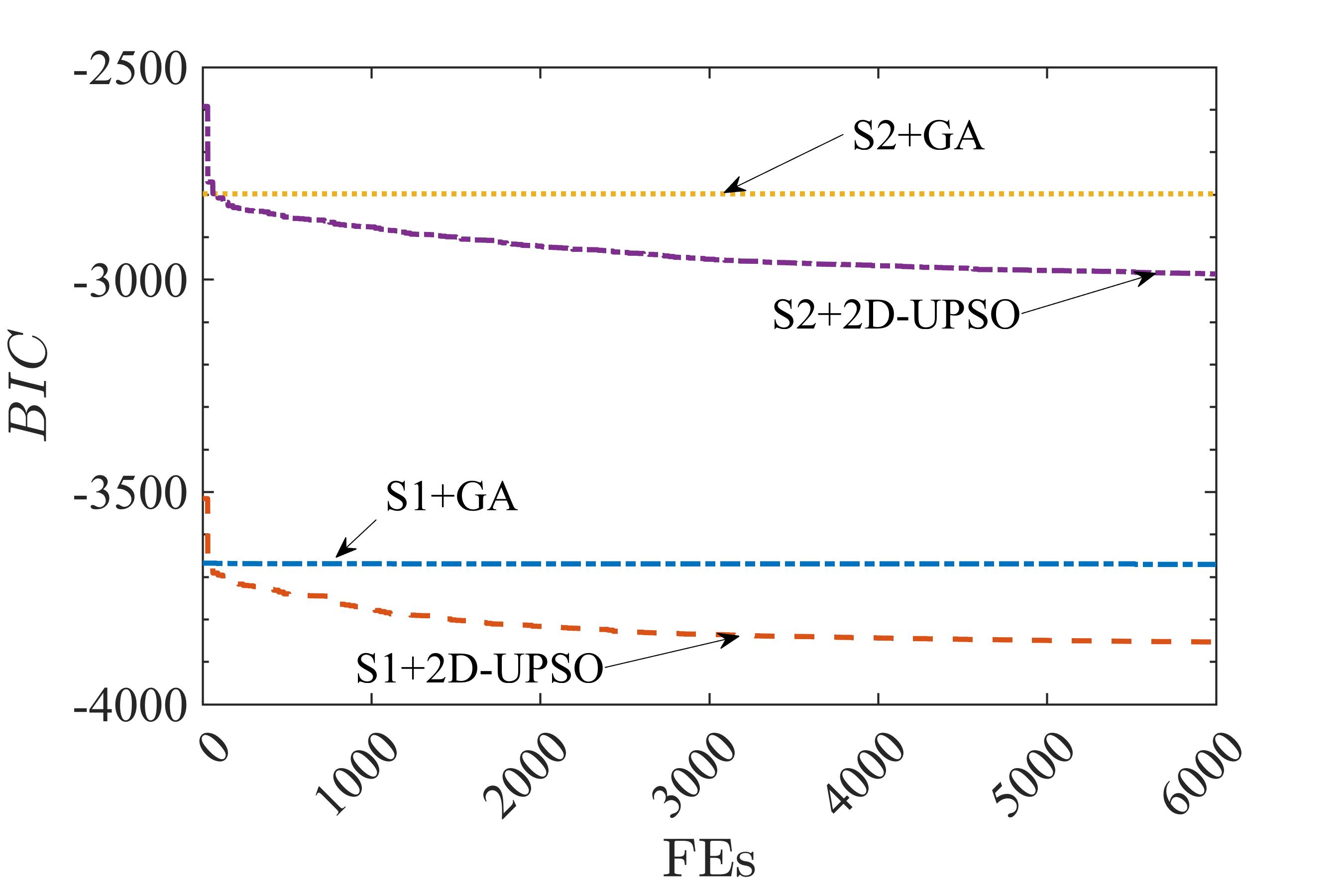}
  \caption{$S1$ and $S2$}
  \label{f:conv1}
\end{subfigure}
\begin{subfigure}{.37\textwidth}
  \centering
  \includegraphics[width=\textwidth]{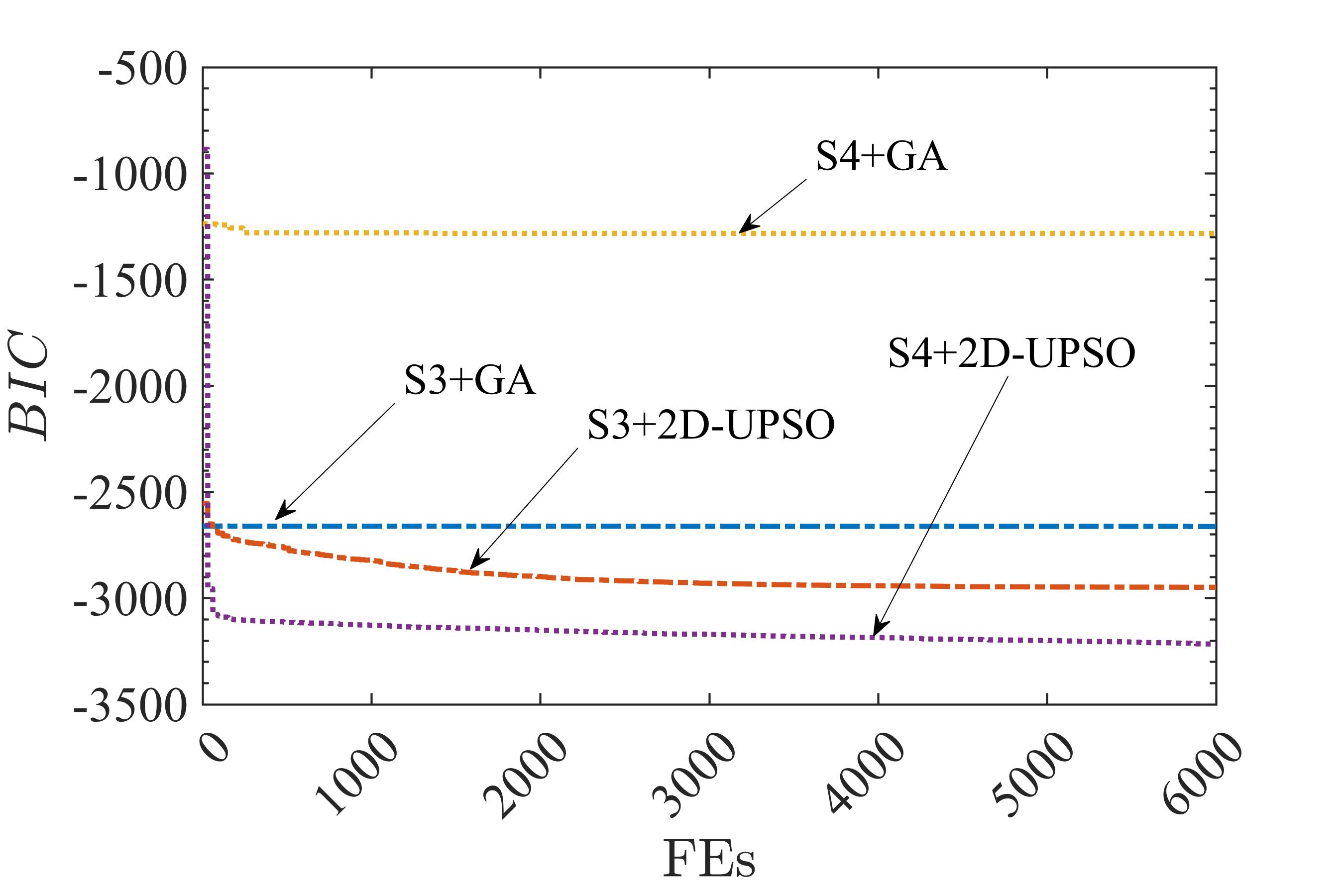}
  \caption{$S3$ and $S4$}
  \label{f:conv2}
\end{subfigure}
\caption{Convergence plots averaged over $40$ runs at SNR$=30dB$}
\label{f:conv}
\end{figure*}
\begin{figure*}[!t]
\centering
\begin{subfigure}{.46\textwidth}
  \centering
  \includegraphics[width=\textwidth]{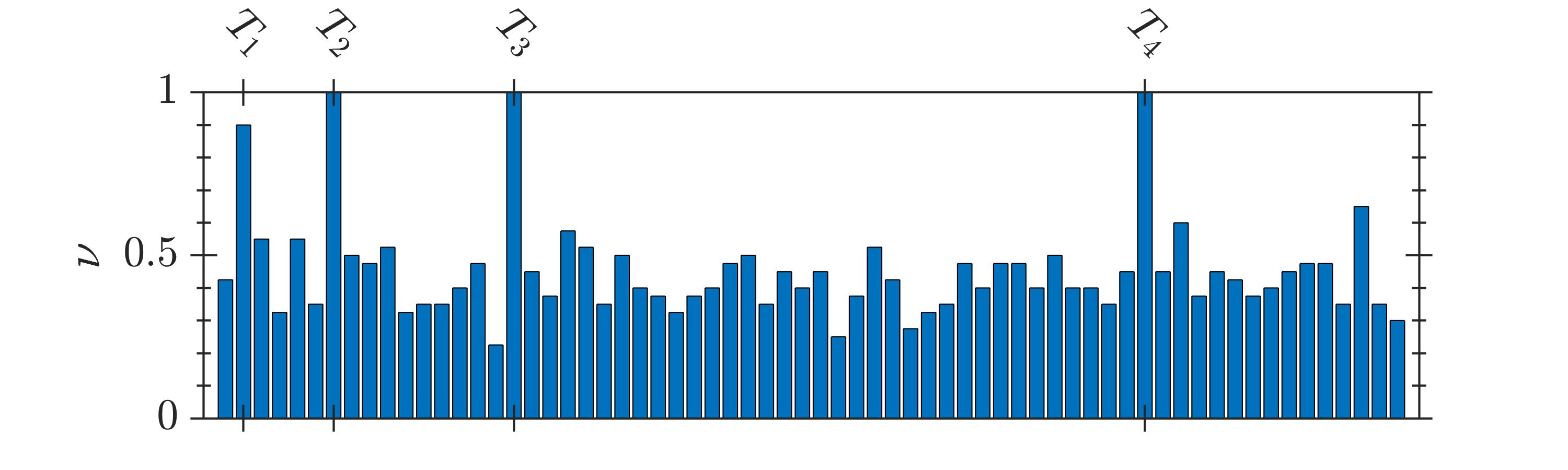}
  \caption{$S1$ + GA}
  \label{f:s11}
\end{subfigure}
\begin{subfigure}{.46\textwidth}
  \centering
  \includegraphics[width=\textwidth]{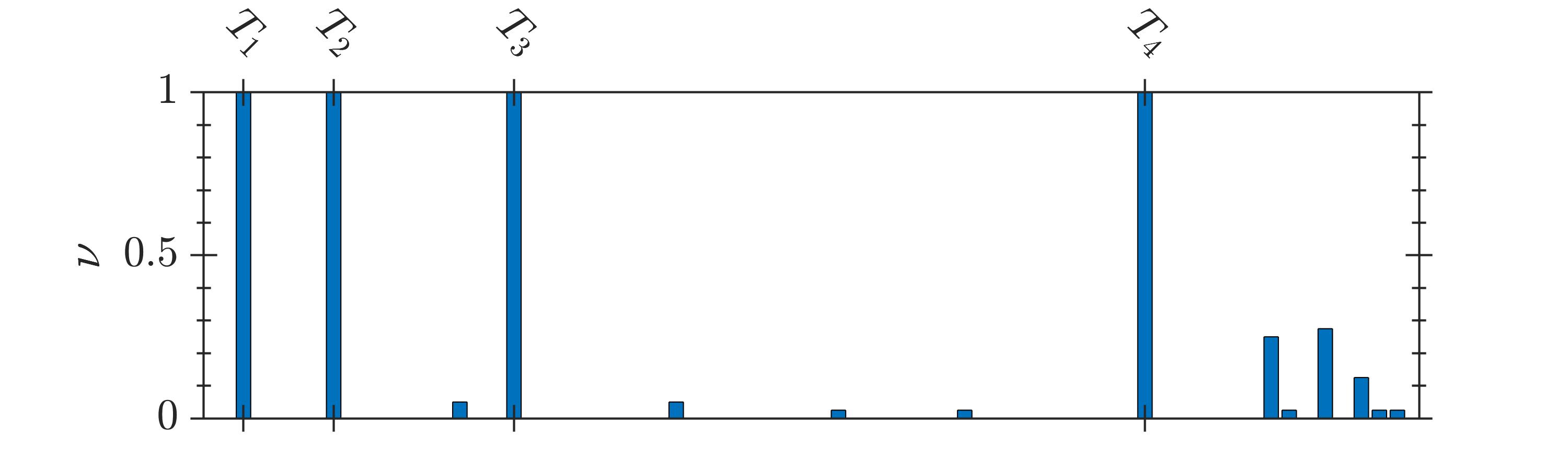}
  \caption{$S1$ + 2D-UPSO}
  \label{f:s12}
\end{subfigure}
\begin{subfigure}{.46\textwidth}
  \centering
  \includegraphics[width=\textwidth]{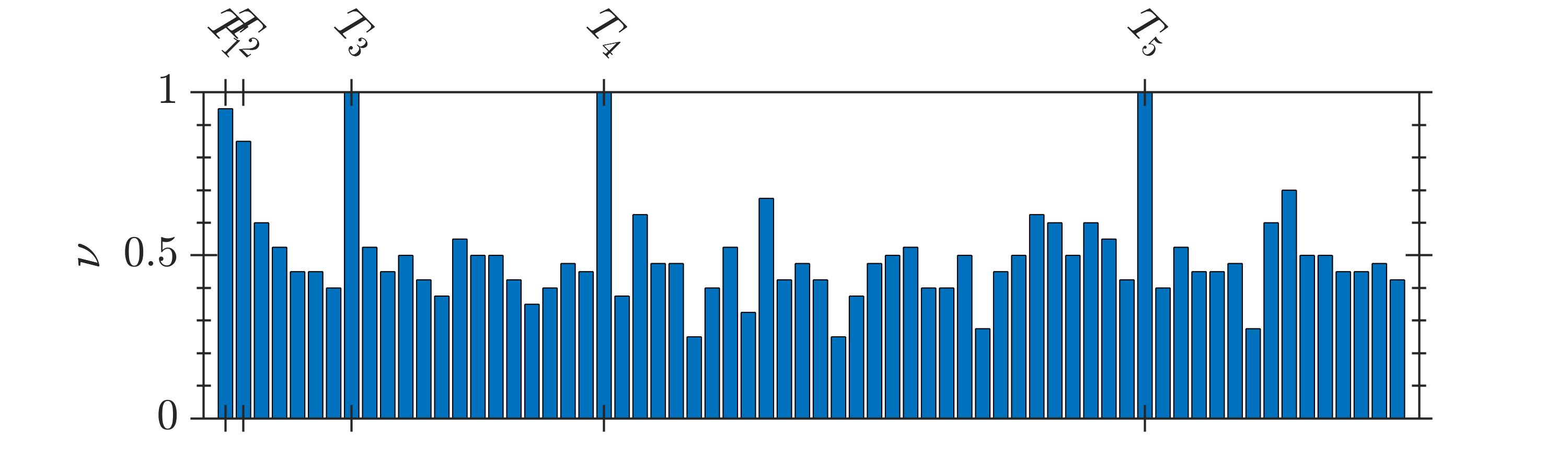}
  \caption{$S2$ + GA}
  \label{f:s21}
\end{subfigure}
\begin{subfigure}{.46\textwidth}
  \centering
  \includegraphics[width=\textwidth]{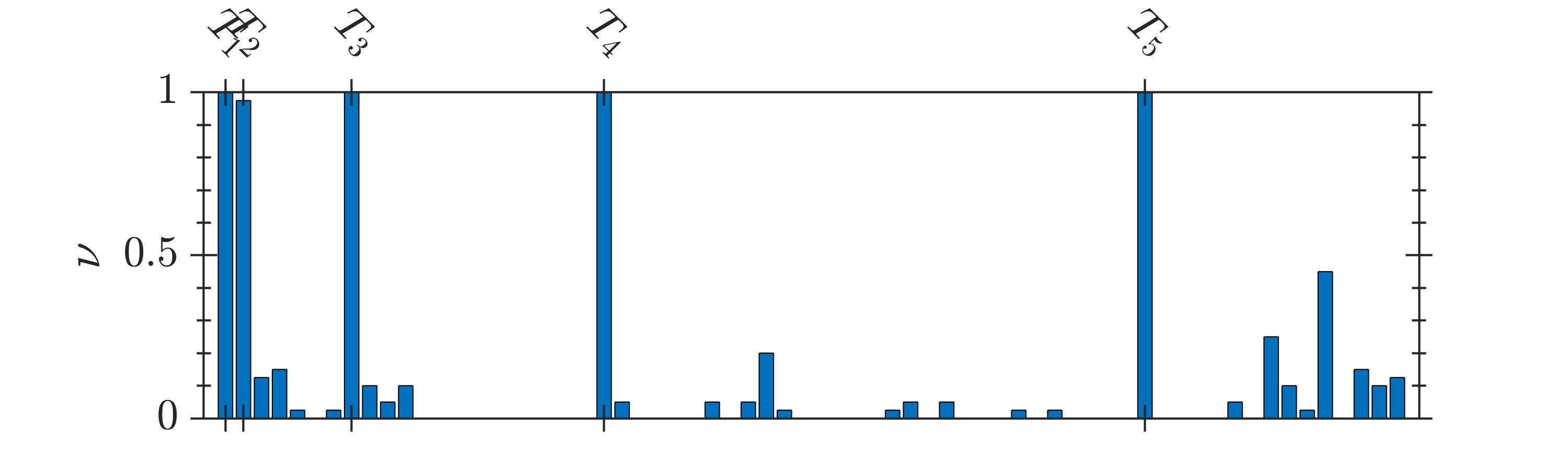}
  \caption{$S2$ + 2D-UPSO}
  \label{f:s22}
\end{subfigure}
\begin{subfigure}{.46\textwidth}
  \centering
  \includegraphics[width=\textwidth]{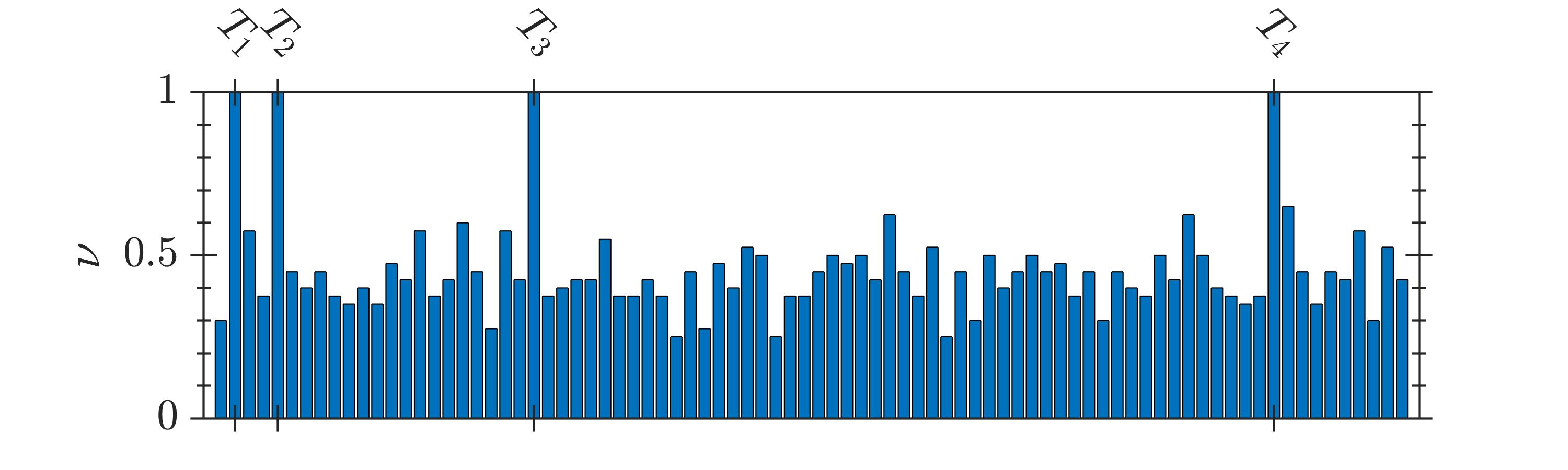}
  \caption{$S3$ + GA}
  \label{f:s31}
\end{subfigure}
\begin{subfigure}{.46\textwidth}
  \centering
  \includegraphics[width=\textwidth]{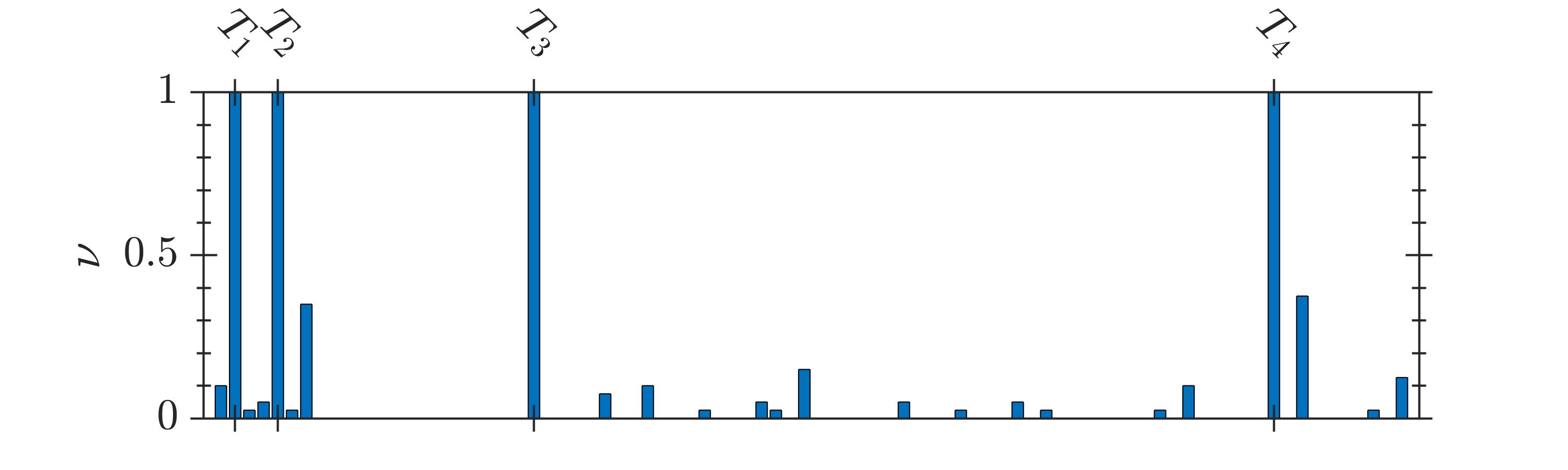}
  \caption{$S3$ + 2D-UPSO}
  \label{f:s32}
\end{subfigure}

\caption{Frequency of the terms identified by GA and 2D-UPSO over 40 independent runs at SNR$=30dB$. The system terms for each system are shown in Appendix~\ref{s:app}. The results for system $S4$ are not shown here due to space constraints.}
\label{f:comp}
\end{figure*}

\subsection{Stage-1 : Choice of the fitness function}
\label{s:r1}
Following the procedure described in Section-IV, the effects of fitness function on structure selection in the presence of various levels of noise was investigated as follows:  

The search performance of 2D-UPSO was recorded for each combination of test system and fitness function in the presence of $3$ distinct noise levels in the input-output data. For a given system, 40 independent runs of 2D-UPSO were carried out for each combination of fitness function (\textit{i.e.} AIC or BIC) and noise level. The objective here is to examine the change in the selection frequency of system and spurious terms with different fitness function. For this purpose, the performance metrics outlined in Section~\ref{s:pm} are determined. Note that, AIC requires the selection of constant $\varrho$ (as seen in~\ref{eq:aic}). To explore the effects of $\varrho$, four different value of $\varrho$ have been used.

The results of this test are shown in Table~\ref{t:fit1}-\ref{t:fit4}, which indicate that in most of the cases the system terms are identified with the frequency $\nu \geq 0.95$, irrespective of the fitness function. However, the fitness function has a significant impact on the spurious terms. For example, for system $S1$, (Table~\ref{t:fit1}), the ratio and maximum frequency of the spurious terms are reduced significantly  from $[0.33, 0.53]$ to $[0.06,0.10]$ when $\varrho$ in AIC is changed from $2$ to $2^8$. Similar effects are observed for system $S2$ (change in $[r,\nu_{max}]$ from $[0.53,0.83]$ with $\varrho=2$ to $[0.08,0.10]$ with $\varrho=2^8$, Table~\ref{t:fit2}) and $S3$, (change in $[r,\nu_{max}]$ from $[0.64,0.90]$ with $\varrho=2$ to $[0.05,0.13]$ with $\varrho=2^8$, Table~\ref{t:fit3}). These observations further underline the role of the fitness function.

Note that the structure selection is essentially a multi-objective problem and the aim of both information criteria, AIC and BIC, is to balance search objectives, \textit{i.e.}, reduction in $e$ and cardinality of the model. The selection of $\varrho$ in (\ref{eq:aic}) is equivalent to guiding a search towards a particular region of the Pareto front. Although this gives the freedom to decision maker to guide the search as per requirement, it is quite challenging to properly tune $\varrho$.

This is evident from the results shown in Table~\ref{t:fit1}-\ref{t:fit4}. These results indicate that $\varrho=2^8$ is suitable for systems $S1$, $S2$ and $S3$, as this value of $\varrho$ leads to significant reduction in the spurious terms (Table~\ref{t:fit1}-\ref{t:fit3}). In contrast, for system $S4$, even the system terms are not identified with $\varrho=2^8$ whereas desirable results are obtained with $\varrho=2$, as seen in Table~\ref{t:fit4}. This can be explained as follows: a higher value of $\varrho$ assigns a high penalty to model structures with higher cardinality (as  seen in (\ref{eq:aic})) and hence drives the search toward a smaller structures. 

From these observations, it is apparent that a proper value of $\varrho$ depends on the structure of the system, \textit{i.e.}, a higher $\varrho$ is effective for the systems with smaller number of terms/cardinality while a lower $\varrho$ is desirable for the system with higher cardinality structures. Given that in practice only the input-output measurement data pairs , $(u,v)$, are available and the cardinality of the structure is not known, it is quite challenging to judiciously select $\varrho$. 

Further, it is interesting to note that, with BIC, all the system terms are identified with $\nu \geq 0.95$ irrespective of the structure cardinality of the corresponding systems, as seen in Table~\ref{t:fit1}-\ref{t:fit4}. In addition the maximum selection frequency of the spurious terms, $\nu_{max}$, is always less than minimum selection frequency of the system terms. Consequently, the spurious terms can easily be removed from the final model structure through a simple thresholding. For this reason, BIC is preferred as the fitness function for the structure selection problem.   

\begin{table*}[!t]
  \centering 
  \small
  \caption{Comparative Evaluation on $S4$ : GA vs. 2D-UPSO $^\dagger$} 
  \label{t:comp}%
   \begin{adjustbox}{max width=0.999\textwidth}
    \begin{threeparttable}
    
    \begin{tabular}{ccccccccccccccccccccccccccc}
    \toprule
     \multirow{2}{*}{\textbf{Algorithm}} & \multirow{2}{*}{\textbf{SNR}} & \multicolumn{23}{c}{\textbf{Selection Frequency (\boldmath$\nu$) of System Terms (\boldmath$T_{system}$)$^\ddagger$}} & \multicolumn{2}{c}{\textbf{Spurious Terms}} \\
    \cmidrule{3-27} & & \boldmath$T_{1}$ & \boldmath$T_{2}$ & \boldmath$T_{3}$ & \boldmath$T_{4}$ & \boldmath$T_{5}$ & \boldmath$T_{6}$ & \boldmath$T_{7}$ & \boldmath$T_{8}$ & \boldmath$T_{9}$ & \boldmath$T_{10}$ & \boldmath$T_{11}$ & \boldmath$T_{12}$ & \boldmath$T_{13}$ & \boldmath$T_{14}$ & \boldmath$T_{15}$ & \boldmath$T_{16}$ & \boldmath$T_{17}$ & \boldmath$T_{18}$ & \boldmath$T_{19}$  & \boldmath$T_{20}$ & \boldmath$T_{21}$ & \boldmath$T_{22}$ & \boldmath$T_{23}$ & \boldmath$r$ & \boldmath$\nu_{max}$ \\
    \midrule
    
    \multirow{3}[2]{*}{\textbf{2D-UPSO}} & $50dB$ & $1$ & $1$ & $1$ & $1$ & $1$ & $1$ & $1$ & $1$ & $1$ & $1$ & $1$ & $1$ & $1$ & $1$ & $1$ & $1$ & $1$ & $1$ & $1$ & $1$ & $1$ & $1$ & $1$ & $0.64$  & $0.88$ \\[0.75ex]
          & $40dB$ & $1$     & $1$     & $1$     & $1$     & $1$     & $1$     & $1$     & $1$     & $1$     & $1$     & $1$     & $1$     & $1$     & $1$     & $1$     & $1$     & $1$     & $1$     & $1$     & $1$     & $1$     & $1$     & $1$     & $0.65$  & $0.95$ \\[0.75ex]
          & $30dB$ & $0.98$  & $1$     & $1$     & $1$     & $1$     & $1$     & $1$     & $1$     & $1$     & $1$     & $1$     & $1$     & $1$     & $1$     & $0.98$  & $1$     & $1$     & $1$     & $1$     & $1$     & $1$     & $1$     & $1$     & $0.62$  & $0.90$ \\[0.75ex]
    \cmidrule{1-27}    \multirow{3}[2]{*}{\textbf{GA}} & $50dB$ & $0.60$  & $0.50$  & $0.50$  & $0.53$  & $1$     & $0.58$  & $1$     & $0.55$  & $0.68$  & $0.60$  & $0.50$  & $0.50$  & $0.55$  & $0.48$  & $0.48$  & $0.58$  & $0.78$  & $1$     & $1$     & $0.75$  & $0.55$  & $0.83$  & $0.55$  & $0.65$  & $0.68$ \\ [0.5ex]
          & $40dB$ & $0.58$  & $0.63$  & $0.45$  & $0.63$  & $1$     & $0.50$  & $1$     & $0.60$  & $0.68$  & $0.48$  & $0.58$  & $0.50$  & $0.63$  & $0.53$  & $0.50$  & $0.53$  & $0.70$  & $1$     & $1$     & $0.73$  & $0.63$  & $0.85$  & $0.63$  & $0.65$  & $0.73$ \\[0.75ex]
          & $30dB$ & $0.58$  & $0.50$  & $0.48$  & $0.58$  & $1$     & $0.63$  & $1$     & $0.38$  & $0.55$  & $0.50$  & $0.73$  & $0.35$  & $0.68$  & $0.48$  & $0.43$  & $0.55$  & $0.85$  & $1$     & $1$     & $0.85$  & $0.58$  & $0.95$  & $0.53$  & $0.65$  & $0.63$ \\
    \bottomrule
    \end{tabular}%

    \begin{tablenotes}
    \small
    \item $^{\dagger}$ `$r$' denotes the ratio of number of redundant terms to the total number of terms
    \smallskip
    \item $^\ddagger$ $ \{ T_1, \dots T_{23} \} \in T_{system}$ denote the system terms and are given in Appendix~\ref{s:app}
    \end{tablenotes}
    \end{threeparttable}
 \end{adjustbox}    
\end{table*}%

\subsection{Stage-2 : Comparative Evaluation of Search Algorithms}

To benchmark the search performance of 2D particle swarms, 2D-UPSO and GA were applied to the test non-linear systems shown in Table~\ref{t:sys}. For this test, BIC (\ref{eq:bic}) was used as the fitness function. For all the systems, $40$ independent runs of each algorithm were carried out and the results are shown in Fig.~\ref{f:conv},~\ref{f:comp} and Table~\ref{t:comp}. 

The search performance of GA and 2D-UPSO is captured through run-length distribution graphs shown in Fig.~\ref{f:conv}. These results indicate that the search process in GA quickly stagnates for all the systems which could be explained by the premature convergence. On the contrary, for the same systems, 2D-UPSO could maintain continual improvement in the fitness function, $J(\cdotp)$, throughout the search.

The selection frequency of the terms, $\nu$, identified by GA and 2D-UPSO is shown in Fig.~\ref{f:comp}. It is observed that even-though GA could identify the system terms, a higher number of spurious terms is present for the systems $S1$-$S3$. In contrast, 2D-UPSO could identify the correct structure of each system with comparatively fewer spurious terms. 

The results of comparative evaluation on $S4$ are shown in Table~\ref{t:comp}. For this system, 2D-UPSO could identify all system terms with $\nu \geq 0.98$, irrespective of the noise level. In contrast, GA could not identify the structure of $S4$ as the selection frequency of many system terms is less than $\nu_{max}$, \textit{i.e.}, it is not possible to remove spurious terms with thresholding as their selection frequency is higher than the system terms. For example, at SNR$=50dB$, $\nu_{max}$ is higher than the selection frequency of $15$ (out of $23$) system terms, as seen in Table~\ref{t:comp}. 

\section{Conclusion}
\label{s:con}

A new two-dimensional learning framework (2D-UPSO) for particle swarms has been applied to address the structure selection problem for a class of non-linear systems which are represented by nonlinear auto-regressive with exogenous inputs (NARX) models. One of the important and advantageous feature of the proposed approach is that it explicitly incorporates the knowledge about the cardinality into the search process. The performance of this approach has been compared with the classical GA based structure selection. For the structure selection, two of the well-known information theoretic criteria such as AIC and BIC have been used as the fitness function. Several non-linear systems have been identified through the proposed approach under various levels of measurement noise using both criteria. First, it was established that BIC criteria is often consistent compared to AIC. Further, it has been established that 2D-UPSO could accurately determine the structure of the systems with significantly less number of spurious terms. 

\appendices
\section{System Terms}
\label{s:app}
\begin{itemize}
\small
    \item System $S1$ : $\{ T_1, \dots T_4\} \in T_{system}$, where, $T_1 \rightarrow y(k-1)$, $T_2 \rightarrow u(k-1)$, $T_3 \rightarrow u(k-1) \ast y(k-1)$ and $T_4 \rightarrow u(k-1)^2$
    \item System $S2$ : $\{ T_1, \dots T_5\} \in T_{system}$, where, $T_1 \rightarrow 5$, $T_2 \rightarrow y(k-1)$, $T_3 \rightarrow u(k-2) \ast y(k-1)$, $T_4 \rightarrow y(k-2)^2$ and $T_5 \rightarrow u(k-1)^2$
    \smallskip
    \item System $S3$ : $\{ T_1, \dots T_4\} \in T_{system}$, where, $T_1 \rightarrow y(k-1)$, $T_2 \rightarrow u(k-1)$, $T_3 \rightarrow u(k-1)^2$ and $T_4 \rightarrow u(k-1)^3$
    \smallskip
    \item System $S4$ : $\{ T_1, \dots T_{23}\} \in T_{system}$, where, $T_1 \rightarrow u(k-1)$, $T_2 \rightarrow u(k-2)$, $T_3 \rightarrow u(k-3)$, $T_4 \rightarrow u(k-1)^2$, $T_5 \rightarrow u(k-1) \ast u(k-2)$, $T_6 \rightarrow u(k-1) \ast  u(k-3)$, $T_7 \rightarrow u(k-2)^2$, $T_8 \rightarrow u(k-2) \ast u(k-3)$, $T_9 \rightarrow u(k-3)^2$, $T_{10} \rightarrow y(k-1)$, $T_{11} \rightarrow y(k-2)$, $T_{12} \rightarrow y(k-3)$, $T_{13} \rightarrow y(k-4)$, $T_{14} \rightarrow y(k-1)^2$, $T_{15} \rightarrow y(k-1) \ast y(k-2)$, $T_{16} \rightarrow y(k-1) \ast y(k-3)$, $T_{17} \rightarrow y(k-1) \ast y(k-4)$, $T_{18} \rightarrow y(k-2)^2$, $T_{19} \rightarrow y(k-2) \ast y(k-3)$, $T_{20} \rightarrow y(k-2) \ast y(k-4)$, $T_{21} \rightarrow y(k-3)^2$, $T_{22} \rightarrow y(k-3) \ast y(k-4)$, $T_{23} \rightarrow y(k-4)^2$
\end{itemize}

\bibliographystyle{IEEEtran}

\end{document}